\documentclass[american,twocolumn,amsmath,amssymb,superscriptaddress]{revtex4-2}
\usepackage[T1]{fontenc}
\usepackage[latin9]{inputenc}
\usepackage{color}
\usepackage{babel}
\usepackage{prettyref}
\usepackage{mathtools}
\usepackage{bm}
\usepackage{amsmath}
\usepackage{amssymb}
\usepackage{graphicx}
\usepackage[unicode=true,pdfusetitle,
 bookmarks=true,bookmarksnumbered=false,bookmarksopen=false,
 breaklinks=false,pdfborder={0 0 1},backref=false,colorlinks=true]
 {hyperref}
\hypersetup{
 linkcolor=blue,citecolor=blue,urlcolor=blue,filecolor=blue}

\makeatletter
\usepackage{babel}
\usepackage{bbm}
\usepackage{MnSymbol}
\usepackage{prettyref}

\usepackage{cprotect}
\usepackage{placeins}

\newrefformat{cap}{\hyperref[#1]{Figure~\ref{#1}}}
\newrefformat{fig}{\hyperref[#1]{Figure~\ref{#1}}}
\newrefformat{tab}{\hyperref[#1]{Table ~\ref{#1}}}
\newrefformat{sec}{\hyperref[#1]{Section~\ref{#1}}}
\newrefformat{subsec}{\hyperref[#1]{Section~\ref{#1}}}
\newrefformat{sub}{\hyperref[#1]{Section~\ref{#1}}}
\newrefformat{cha}{\hyperref[#1]{Chapter~\ref{#1}}}
\newrefformat{app}{\hyperref[#1]{Appendix~\ref{#1}}}

\makeatother

\begin{document}
\global\long\def\l{\langle}
\global\long\def\T{\mathrm{T}}
\global\long\def\r{\rangle}
\global\long\def\w{\omega}
\global\long\def\ll{\llangle}
\global\long\def\rr{\rrangle}
\global\long\def\tr{\mathrm{tr}}
\global\long\def\tx{\tilde{x}}
\global\long\def\hj{\hat{j}}
\global\long\def\jb{\bar{j}}
\global\long\def\gb{\bar{g}}
\global\long\def\e{\exp}
\global\long\def\t{\tau}
\global\long\def\N{\mathcal{N}}
\global\long\def\hp{\hat{\phi}}
\global\long\def\i{\mathfrak{i}}
\global\long\def\kj{\mathbf{k}_{j}}
\global\long\def\extr{\text{extr}}
\global\long\def\erf{\text{erf}}

\title{Linking Network- and Neuron-Level Correlations by Renormalized Field
Theory}
\author{Michael Dick$^{*}$}
\affiliation{Institute of Neuroscience and Medicine (INM-6) and Institute for Advanced
Simulation (IAS-6) and JARA-Institute Brain Structure-Function Relationships
(INM-10), Jülich Research Centre, Jülich, Germany}
\affiliation{Department of Computer Science 3 - Software Engineering, RWTH Aachen
University, Aachen, Germany}
\affiliation{Peter Grünberg Institut (PGI-1) and Institute for Advanced Simulation
(IAS-1), Jülich Research Centre, Jülich, Germany}
\email{mi.dick@fz-juelich.de}

\author{Alexander van Meegen}
\affiliation{Institute of Neuroscience and Medicine (INM-6) and Institute for Advanced
Simulation (IAS-6) and JARA-Institute Brain Structure-Function Relationships
(INM-10), Jülich Research Centre, Jülich, Germany}
\affiliation{Institute of Zoology, University of Cologne, 50674 Cologne, Germany}
\author{Moritz Helias}
\affiliation{Institute of Neuroscience and Medicine (INM-6) and Institute for Advanced
Simulation (IAS-6) and JARA-Institute Brain Structure-Function Relationships
(INM-10), Jülich Research Centre, Jülich, Germany}
\affiliation{Department of Physics, Faculty 1, RWTH Aachen University, Aachen,
Germany}
\date{\today}
\begin{abstract}
It is frequently hypothesized that cortical networks operate close
to a critical point. Advantages of criticality include rich dynamics
well-suited for computation and critical slowing down, which may offer
a mechanism for dynamic memory. However, mean-field approximations,
while versatile and popular, inherently neglect the fluctuations responsible
for such critical dynamics. Thus, a renormalized theory is necessary.
We consider the Sompolinsky-Crisanti-Sommers model which displays
a well studied chaotic as well as a magnetic transition. Based on
the analogue of a quantum effective action, we derive self-consistency
equations for the first two renormalized Greens functions. Their self-consistent
solution reveals a coupling between the population level activity
and single neuron heterogeneity. The quantitative theory explains
the population autocorrelation function, the single-unit autocorrelation
function with its multiple temporal scales, and cross correlations.
\end{abstract}
\maketitle

\section{Introduction}

\subsection{Critical Neural Dynamics}

\label{sec:networks_and_transitions}

Both experiments and models of cortical networks suggest that the
brain is operating close to a phase transition \citet{Beggs04_5216,Chialvo10_744,Priesemann14_80,Fontenele19_208101}.
Indicators for this phenomenon are for example found in parallel recordings
of neuronal cell cultures for which the number of coactive neurons
shows power law distributions \citep{Beggs04_5216}. The pattern of
neuronal activity, in this case referred to as an avalanche, looks
identical on several length and time scales, which suggests a continuous
phase transition \citep{Goldenfeld92}. The transition point of a
continuous phase transition is synonymous with fluctuations on all
time scales dominating the system's behavior. This makes it difficult
to obtain systematic approximations, rendering continuous phase transitions
notoriously hard to treat.

More recent work \citep{Priesemann14_80} suggests that the measured
critical behavior could be due to the inherent sub-sampling in neuronal
recordings which are so far only able to capture a fraction of a network's
neurons. Even though this work shows that the observed critical exponents
are influenced through measurement effects, it still suggests that
the system is slightly sub-critical. Closeness to such a transition
comes with numerous benefits. Critical slowing down, the effect of
increasing and, at the transition point, even diverging decay constants
makes a large spectrum of time constants available to the network.
This leads to optimal memory capacity as has been shown using stochastic
artificial neuronal networks \citep{Toyoizumi11_051908,Schuecker18_041029},
and maximal computational performance \citep{Legenstein07_323}.

So far it is unclear what phase transition the brain operates close
to. However, there are two popular candidates: The first is a transition
into a chaotic regime, meaning that infinitesimal changes in the neuron
dynamics are progressively amplified \citep{Kadmon15_041030,Dahmen19_13051,Sompolinsky88_259}.
The other is known as avalanche-like criticality \citep{Priesemann14_80,Beggs03_11167}.
Avalanches can be viewed through the lens of branching processes,
treating the propagation of neuronal activity as the children and
further descendants of a spontaneously emitted spike. Below criticality,
each spike has on average less than one child, leading to activity
being driven by external input and a quick decay of all child processes.
Above criticality each neuron is on average responsible for more than
one spike, leading to escalating activity. At the critical point itself,
where there is on average one child spike, long transients are possible
and complex behavior can emerge.

Both transitions are well studied in isolation in different models,
making direct comparisons difficult. For both, the transition to chaos
and avalanches, models exist which show critical behavior, but there
has not yet been a study of a model supporting both phase transitions.

\subsection{Model and Renormalized Theory}

We want to pave the way to a comparison of the two phase transitions
in this paper. To this end we focus on an adaptation of the popular
and simple model by Sompolinsky, Crisanti, and Sommers \citep{Sompolinsky88_259}.
It models the activity of $N$ neurons in a randomly connected recurrent
neural network. The activity of a single neuron $i$ is denoted by
$x_{i}(t)$ and it is governed by the coupled system of stochastic
differential equations
\begin{align}
\dot{x}_{i}+x_{i} & =\sum_{j=1}^{N}J_{ij}\phi_{j}+\xi_{i},\label{eq:diffeq_motion}
\end{align}
where we use the abbreviation $\phi_{i}\equiv\phi(x_{i})$ and the
driving noise $\xi_{i}$ is assumed to be Gaussian white noise with
zero mean and covariance $\langle\xi_{i}(t)\xi_{j}(s)\rangle=D\,\delta_{ij}\,\delta(t-s)$.
Here $\phi$ is an arbitrary activation function for most of this
paper; in simulations we chose an error function $\phi(x)=\erf(\frac{\sqrt{\pi}}{2}x)=\int_{0}^{x}\,e^{-\frac{\pi}{4}z^{2}}\,dz$
where the scaling ensures that the slope at the origin is unity. In
the absence of the right hand side in \eqref{eq:diffeq_motion}, the
activity decays exponentially with unit time constant. The right hand
side represents the input to the neuron. The first part comes from
all other neurons determined via the activation function $\phi$ and
the connectivity $J$, whose $N\times N$ weights are distributed
according to a Gaussian with mean $\bar{g}/N$ (which is often set
to zero) and variance $g^{2}/N$. We will refer to $\bar{g}$ as the
mean and $g^{2}$ as the variance of the connectivity as the factor
$N^{-1}$ is simply chosen such that mean and fluctuations of the
input to a neuron do not scale with the total number of neurons. The
second source of input is a random white-noise $\xi_{i}$ with noise
intensity $D$ modeling external input from other brain areas.

To link this model to the two forms of criticality mentioned above,
we consider $\bar{g}$ as the control parameter for avalanche-like
activity; if non-zero and positive it controls the strength by which
the average population activity at a certain instant excites and maintains
the activity at the next point in time. More formally, in the limit
of large $N$, the parameter $\bar{g}$ controls a single real outlier
eigenvalue of the connectivity matrix, $\bar{\lambda}=\bar{g}$ \citep{Tao2011_231,Schuessler20_013111},
with corresponding eigenvector $(1,\ldots,1)$. The latter is a mode
in which all neurons act in unison, a cartoon of what happens in a
neuronal avalanche. If this eigenvalue $\bar{\lambda}$ crosses unity,
the silent fixed point of the noiseless ($D=0$) model becomes unstable
in this very direction \citep{Mastrogiuseppe18_609}. The transition
to chaos, in contrast, is predominantly controlled by the parameter
$g$. Studying the eigenvalues of the connectivity, $g$ controls
the radius of the bulk of eigenvalues which are uniformly distributed
in a circle with radius $g$ around the origin of the complex plane.
Again, a critical point is reached if this radius reaches unity, because
then all eigenmodes with $\Re(\lambda_{i})\simeq1$ show critically
slow dynamics. In the noiseless case $D=0$ (and for $\bar{g}=0$)
this points marks the onset of chaotic dynamics \citep{Sompolinsky88_259}.

\begin{figure}[h]
\includegraphics[width=1\columnwidth]{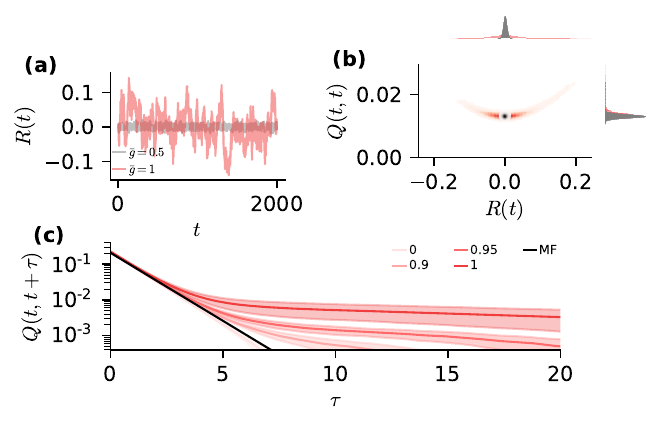}\caption{\textbf{(a)} Population-averaged activity $R(t)$ for $\bar{g}=0.5$
(gray) and $\bar{g}=1$ (red). \textbf{(b)} Auxiliary fields $Q$
and $R$, proportional to population-averaged output autocorrelation
and activity, respectively, binned for each point in time. \textbf{(c)}
Time-lagged, population-averaged, stationary autocorrelation $Q(t,t+\tau)$
simulated for different values of $\bar{g}$ (shades of red) and mean
field prediction (black) plotted logarithmically. Remaining network
parameters: $\phi(x)=\protect\erf(\frac{\sqrt{\pi}}{2}x)$, $N=1000$,
$g=0.5$, and $D=0.1$.}
\label{fig:R}
\end{figure}

The theoretical approach to the disordered system described by \eqref{eq:diffeq_motion}
is based on mean-field approximations on auxiliary fields like 
\begin{align}
R(t):= & \frac{\bar{g}}{N}\sum_{j=1}^{N}\phi_{j}(t),\label{eq:def_R}\\
Q(s,t):= & \frac{g^{2}}{N}\sum_{j=1}^{N}\phi_{j}(s)\phi_{j}(t),\label{eq:def_Q}
\end{align}
since they give a way to obtain an effective low-dimensional set of
equations describing the collective behavior. This approach has been
used to show a transition to chaos with $g^{2}$ acting as control
parameter \citep{Sompolinsky88_259}, which has been studied extensively
\citep{Kadmon15_041030,Mastrogiuseppe17_e1005498}. As discussed above,
the mean $\bar{g}$ of the connectivity can also take the form of
a control parameter \citep{Mastrogiuseppe18_609}: as seen in \prettyref{fig:R}a
the network exposes large fluctuations in its population-averaged
activity as $\bar{g}$ approaches the transition point given by $\bar{g}$
close to unity for $g<1$. Close to this criticality the fluctuations
of $R$ will influence $Q$ as can be seen in \prettyref{fig:R}b,
leading on average to a larger autocorrelation. In this work we derive
an analytical way to analyze the network's behavior close to this
transition using $\bar{g}$ as a control parameter, taking into account
the fluctuations of population-averaged activity and its effect on
the autocorrelation function. 

The proper treatment of fluctuations comes with some technical difficulties.
Mean-field approaches, albeit being very popular in the field \citep{Kadmon15_041030,Vreeswijk96_5522,Amit97,Brunel00},
neglect fluctuations of the auxiliary fields. This effect can be seen
in \prettyref{fig:R}c, which shows the population averaged autocorrelation
simulated for several values of $\bar{g}$ close to unity compared
to the analytical mean-field solution, which corresponds in this case
to a network with $\bar{g}=0$. One clearly sees that the mean-field
results (black) are not sufficient to describe the second time constant,
which grows with rising $\bar{g}$ (plotted in shades of red).

One way of taking these fluctuations into account is by means of Legendre
transformation methods \citep{Vasiliev98}. These provide a way to
derive a set of self-consistent equations that resum these fluctuations
and are therefore able to describe the observed behavior.

\subsection{Outline}

We will derive a set of self-consistent equations for the mean and
the fluctuations of the auxiliary fields \eqref{eq:def_R} and \eqref{eq:def_Q}.
Such self-consistent schemes are commonplace in other fields of physics
\citep{Vasiliev98,Berges04_0409233}. These approximations are typically
formulated in the language of a field theory. As a first step, we
therefore formulate the dynamical equations in this language. Initially
we will leave the activation function $\phi$ general; all we ask
of it is to vanish at zero and to possess a Fourier transform. This
set of self-consistency equations in particular exposes how the fluctuations
of the population-averaged activity $R$ influence the population-averaged
autocorrelation $Q$, as shown empirically in \prettyref{fig:R}b
and \prettyref{fig:R}c. The theory also allows us to compute pairwise
correlations averaged across all pairs of neurons in the network.
Lastly, the theory proposes that stimulations that excite the population-averaged
activity $R$ also influence the heterogeneity of the response across
neurons, as measured by $Q$.

\section{Self-consistent second-order statistics}

\subsection{Action for Auxiliary Fields}

First, we translate~\eqref{eq:diffeq_motion} into the language of
field theory. To this end, it is instructive to first look at the
noise expectation value of an operator $G[\bm{x}]$ constrained to
the dynamics of~\eqref{eq:diffeq_motion}. This can be achieved with
help of the Martin-Siggia-Rose-de~Dominicis-Janssen formalism \citep{Martin73,janssen1976_377,dedominicis1976_247}
(for pedagogic reviews see \citep{Chow15,Hertz16_033001,Helias20_970})
and results in

\begin{align}
\l G[\bm{x}]\r_{\bm{x}|\bm{J}} & =\int_{\bm{x}}\,\l\delta[\bm{\dot{x}}+\bm{x}-\bm{J}\phi(\bm{x})-\bm{\xi}]\r_{\bm{\xi}}\,G[\bm{x}]\nonumber \\
 & =\int_{\bm{x},\bm{\tilde{x}}}\,e^{S_{0}[\bm{x},\bm{\tilde{x}}]-\bm{\tilde{x}}^{T}\bm{J}\phi(\bm{x})}\,G[\bm{x}].\label{eq:avg_G}
\end{align}
Here, $\int_{\bm{x}}$ denotes an integral over the trajectories
of all neurons and we used $\delta(x)=\tfrac{1}{2\pi i}\int_{-i\infty}^{i\infty}e^{\tilde{x}x}\,d\tilde{x}$
for every time step and neuron and defined the action
\begin{align}
S_{0}[\bm{x},\bm{\tilde{x}}] & :=\bm{\tx}^{\T}\left(\partial_{t}+1\right)\bm{x}+\frac{D}{2}\bm{\tx}^{\T}\bm{\tx}\label{eq:def_S0}
\end{align}
with the short hand notations $\bm{a}^{\T}\bm{b}=\sum_{i=1}^{N}\int_{0}^{t}ds\,a_{i}(s)b_{i}(s)$
and $\bm{a}^{\T}\bm{M}\bm{b}=\sum_{i,j=1}^{N}\int_{0}^{t}ds\,a_{i}(s)M_{ij}b_{j}(s)$.

This allows the definition of a characteristic functional $Z[\bm{l}]$
by setting $G[\bm{x}]=\exp(\bm{l}^{\T}\bm{x})$. The source $\bm{l}$
in the exponent allows us to take derivatives which in turn yield
properly normalized moments after evaluating at the physical value
$\bm{l}=0$ of the sources. These sources need not be linear in $\bm{x}$
and could even couple to entirely different quantities. Until we need
them we will leave them out and first consider only the partition
function.

Eventually, we are interested in self averaging quantities like the
mean \eqref{eq:def_R} and the autocorrelation function \eqref{eq:def_Q};
thus, we further average over realizations of the connectivity $J_{ij}\stackrel{\text{i.i.d.}}{\sim}\N(\bar{g}/N,g^{2}/N)$
which only affects the term $-\bm{\tilde{x}}^{\T}\bm{J}\phi(\bm{x})$
and yields

\begin{align}
\l e^{-\bm{\tilde{x}}^{\T}\bm{J}\phi(\bm{x})}\r_{\bm{J}} & =\int_{y}\exp\bigg(\frac{N}{2}\,y^{\T}Ky+\sum_{i=1}^{N}y^{\T}f[z_{i}]\bigg).
\end{align}
Here, we introduced the population-averaged auxiliary fields $R$
defined in \eqref{eq:def_R} and $Q$ defined in \eqref{eq:def_Q}
via Hubbard-Stratonovich transformations and their respective response
fields $\tilde{R}$ and $\tilde{Q}$ analogously to the introduction
of $\bm{\tx}$. Furthermore, we introduced several short-hand notations:
First, we denote $\bm{x}$ and $\bm{\tx}$ in combination as $\bm{z}=(\bm{x},\bm{\tx})$
and $R$, $\tilde{R}$, $Q$, and $\tilde{Q}$ in combination as $y=(R,\tilde{R},Q,\tilde{Q})$.
Second, we abbreviate $y^{\T}f[z_{i}]=-\tx_{i}^{\T}R-\bar{g}\phi_{i}^{\T}\tilde{R}+\frac{1}{2}\tx_{i}^{\T}Q\tx_{i}-g^{2}\phi_{i}^{\T}\tilde{Q}\phi_{i}$.
Third, we define $K=\left(\begin{array}{cc}
\sigma_{x} & 0\\
0 & \sigma_{x}
\end{array}\right)$ where $\sigma_{x}=\left(\begin{array}{cc}
0 & 1\\
1 & 0
\end{array}\right)$, leading to $\frac{1}{2}\,y^{\T}Ky=\tilde{R}^{\T}R+\tilde{Q}^{\T}Q$.
In summary, the introduced notation allow us to write
\begin{align*}
\langle\l G(\bm{x})\r_{\bm{x}|\bm{J}}\rangle_{\bm{J}}= & \int_{y}e^{\frac{N}{2}\,y^{\T}Ky}\prod_{i=1}^{N}\int_{z_{i}}e^{S_{0}[z_{i}]+y^{\T}f[z_{i}]}\,G(x_{i})
\end{align*}
for any factorizing $G(\bm{x})=\prod_{i=1}^{N}G(x_{i})$.

We see that the part of the partition function that describes individual
neurons factorizes into $N$ identical factors. This leaves a partition
function for the four auxiliary fields interacting with a single neuron
\begin{align*}
\int_{y}e^{\frac{N}{2}\,y^{\T}Ky}\prod_{i=1}^{N}\int_{z_{i}}e^{S_{0}[z_{i}]+y^{\T}f[z_{i}]} & =\int_{y}\e\left(N\,S[y]\right),
\end{align*}
where we defined the action for the auxiliary fields as
\begin{align}
S[y] & :=\frac{1}{2}y^{\T}Ky+\mathcal{W}[y],\label{eq:action_y}\\
\mathcal{W}[y] & :=\ln\int_{z}\e\left(S_{0}[z]+y^{\T}f[z]\right),\label{eq:def_W_cal_y}
\end{align}
reducing the dimensionality of the problem from $N$ neurons to the
six fields $y$ and $z$. We note that $\mathcal{W}[y]$ has the form
of a cumulant-generating functional for $f[z]$.

\subsection{Mean-Field Phase Diagram}

As the lowest order (mean-field) approximation one can treat the path
integrals $\int_{y}$ in saddle point approximation, replacing the
auxiliary fields with their most likely values obtained from the condition
$\delta S[y]/\delta y_{i}\stackrel{!}{=}0$, which yields \citep{Mastrogiuseppe17_e1005498,Mastrogiuseppe18_609,Schuecker18_041029,Helias20_970}
\begin{align*}
y^{\ast} & =(R^{\ast},\tilde{R}^{\ast},Q^{\ast},\tilde{Q}^{\ast})\\
 & =(\bar{g}\mu_{\phi},0,g^{2}C_{\phi\phi},0),
\end{align*}
with
\begin{align*}
\mu_{\phi}(t) & =\langle\phi(t)\rangle,\\
C_{\phi\phi}(t,s) & =\langle\phi^{2}(s,t)\rangle,
\end{align*}
where $\langle\ldots\rangle$ is the measure determined by the action
\eqref{eq:action_y} and $\phi^{2}(s,t):=\phi(t)\phi(s)$.

We are now left with path integral for a single neuron and its response
field which corresponds to the stochastic differential equation
\begin{align}
\dot{x}+x & =\xi+\eta,\label{eq:scs_dmft}
\end{align}
where $\eta$ is a Gaussian Process with
\begin{align}
\ll\eta(t)\rr & =\bar{g}\mu_{\phi}(t),\label{eq:eta_mean}\\
\ll\eta(s)\eta(t)\rr & =g^{2}C_{\phi\phi}(s,t),\label{eq:eta_corr}
\end{align}
where we use $\ll\ldots\rr$ to denote cumulants (connected correlation
functions). For an error function as the nonlinearity these expectations
can be calculated analytically in terms of statistics of the neuron
activity \citep{vanMeegen21_043077}. Thus \eqref{eq:scs_dmft} can
be solved efficiently in a self-consistent manner.

\begin{figure}
\includegraphics[width=1\columnwidth]{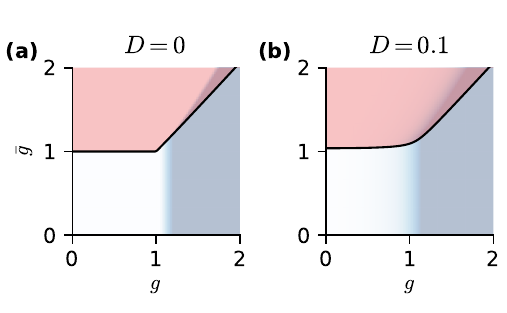}

\caption{Mean-field phase diagram spanned by $\bar{g}$ and $g$ for (\textbf{a})
the noiseless case ($D=0$) and (\textbf{b}) noise-driven dynamics
($D=0.1$). The red shading quantifies the absolute population activity
$|R|$, which is the order parameter for ferromagnetic activity and
the gray shading quantifies the dynamic variability $Q$, which for
$D=0$ is the order parameter indicating the onset of chaotic activity.
The black curves show where these values become nonzero. The dynamic
variability $Q$ does not vanish in the presence of noise.\label{fig:param_space_nonoise}}
\end{figure}

For the case of vanishing noise ($D=0$) the saddle-point approximation
recovers the phase diagram from \citep[Fig. 1B]{Mastrogiuseppe18_609}
(see also \prettyref{fig:param_space_nonoise} a): The system exhibits
a transition from a state with a vanishing order parameter $R=0$
to a state with a broken symmetry where $|R|>0$ at a critical value
$\bar{g}=\bar{g}_{c}$. For the case with noise ($D>0$), the point
of transition in addition depends on the noise amplitude $\bar{g}=\bar{g}_{c}(g,D)$;
see \eqref{eq:D_crit} for an explicit expression for $D=D(\bar{g}_{c},g)$
which can be solved for $\bar{g}_{c}=\bar{g}_{c}(g,D)$.

\subsection{Equations of State to 1-loop Order}

We are especially interested in the transition to structured activity
$|R|>0$ driven by the mean connectivity $\bar{g}$. We expect this
transition to be accompanied by fluctuations of the auxiliary field
\eqref{eq:def_R} and thus aim to derive a description treating population-wide
fluctuations systematically. The population level activity is captured
by the auxiliary fields. It is thus natural to introduce sources for
these fields and for their square to measure their fluctuations. This
leads us to the definition of a moment generating functional

\begin{align}
Z[j,k] & =\int_{y}e^{N\,\mathcal{W}[y]+j^{\T}y+\frac{1}{2}y^{\T}k^{\T}y},\label{eq:Z_SCS_def}
\end{align}
which yields the first and second moment of $y$ upon differentiation
by $j$ and $k$, respectively, at the physically relevant value of
the sources $j=0$ and $k=N\,K$, by comparison to \eqref{eq:action_y}.
Our aim is to obtain self-consistency equations for the first two
moments. It is therefore helpful to define an ensemble where these
two moments are fixed. This is achieved by performing a second-order
Legendre transform to the effective action 
\begin{align*}
\Gamma[\alpha_{1},\alpha_{2}] & =\text{extr}_{j,k}j^{\T}\alpha_{1}+\frac{1}{2}k^{\T}\alpha_{2}-\ln Z[j,k]\\
 & =\text{extr}_{j,k}-\ln\int_{y}e^{N\,\mathcal{W}[y]+j^{\T}(y-\alpha_{1})+\frac{1}{2}k^{\T}(y^{2}-\alpha_{2})},
\end{align*}
which fixes the system's first two moments $\alpha_{1},\alpha_{2}$
of the auxiliary fields $y$; here $k^{\T}y^{2}$ is meant as a bilinear
form in $y$. The equations of state then yield self-consistency equations
\begin{align}
\frac{\delta\Gamma}{\delta\alpha_{1}} & =j=0,\label{eq:eqs_of_state_alpha}\\
\frac{\delta\Gamma}{\delta\alpha_{2}} & =\frac{1}{2}k=\frac{1}{2}N\,K.\nonumber 
\end{align}
Below, we will perform a fluctuation expansion of $\Gamma$. To ensure
that only connected diagrams appear in the expansion \citet{Vasiliev98},
we describe the system via its cumulants $\beta_{1}=\alpha_{1}$ and
$\beta_{2}=\alpha_{2}-\alpha_{1}^{2}$ and define an effective action
in these new coordinates (see \prettyref{app:legendre_in_cumulants})
\begin{align*}
\Gamma[\beta_{1},\beta_{2}] & =\text{extr}_{\hat{j},k}-\ln\,\int_{y}e^{N\,\mathcal{W}[y]+\hat{j}^{\T}(y-\beta_{1})+\frac{1}{2}k^{\T}[(y-\beta_{1})^{2}-\beta_{2}]},
\end{align*}
where $\hat{j}:=j+k\beta_{1}$. Following \citet{Vasiliev98} we here
use the notation of $\alpha_{n}$ for the $n$-th moment and $\beta_{n}$
for the $n$-th cumulant. We thus have $(\beta_{1})_{1}=\ll R\rr=:R^{\ast}$
and $(\beta_{1})_{3}=\ll Q\rr=:Q^{\ast}$. The other two components
of $\beta_{1}$ are zero, as they are cumulants of response fields.
For $\beta_{2}$ we will use the notation $\beta_{ij}=(\beta_{2})_{ij}$
as it comes up frequently. So we have $\beta_{11}$ as the autocorrelation
of $R$, $\beta_{12}$ and $\beta_{21}$ as its response functions
and again $\beta_{22}=0$ as a cumulant of only response fields. The
equations of state \prettyref{eq:eqs_of_state_alpha} in the new coordinates
take the form 
\begin{align}
\frac{\delta\Gamma[\beta_{1},\beta_{2}]}{\delta\beta_{1}} & =j+\beta_{1}k=\beta_{1}NK,\label{eq:first_eq_of_state}\\
\frac{\delta\Gamma[\beta_{1},\beta_{2}]}{\delta\beta_{2}} & =\frac{1}{2}k=\frac{1}{2}NK.\label{eq:second_eq_of_state}
\end{align}
Writing the problem in this way uses the yet unknown fluctuation-corrected
self-consistent values for the first and second-order statistics which
become accessible via the equations of state.

Solving the equations of state is difficult in general but as $S[y]\propto N$
a loop-wise expansion becomes meaningful. Up to one-loop order and
neglecting additive constants we get by expanding $\mathcal{W}[y]=\mathcal{W}[\beta_{1}]+\frac{1}{2}(y-\beta_{1})^{\T}\mathcal{W}^{(2)}[\beta_{1}](y-\beta_{1})$
and performing the resulting Gaussian integral over the fluctuations
$\delta y=y-\beta_{1}$ 
\begin{align*}
\Gamma_{\text{1-loop}}[\beta_{1},\beta_{2}]= & -N\,\mathcal{W}[\beta_{1}]+\frac{1}{2}k^{\T}\beta_{2}\\
 & +\frac{1}{2}\ln\det(-N\mathcal{W}^{(2)}[\beta_{1}]-k).
\end{align*}
Note that the terms linear in the fluctuations do not contribute to
one-loop order. Using the stationarity condition $\frac{\delta}{\delta k}\Gamma_{\text{1-loop}}[\beta_{1},\beta_{2}]=0$,
we obtain $\beta_{2}=(-N\mathcal{W}^{(2)}[\beta_{1}]-k)^{-1}$ which
simplifies the effective action to
\begin{align*}
\Gamma_{\text{1-loop}}[\beta_{1},\beta_{2}]= & -N\,\mathcal{W}[\beta_{1}]-\frac{1}{2}N\,\mathcal{W}^{(2)}[\beta_{1}]^{\T}\beta_{2}\\
 & -\frac{1}{2}\ln\det(\beta_{2}),
\end{align*}
where we suppressed the inconsequential constant $-\frac{1}{2}\tr\,\mathbb{I}.$
Up to one-loop order and evaluated at their true value $j=0$ and
$k=N\,K$ the first equation of state \eqref{eq:first_eq_of_state}
reads
\begin{align}
\frac{\delta\Gamma_{\text{1-loop}}[\beta_{1},\beta_{2}]}{\delta\beta_{1}} & =-N\,\mathcal{W}^{(1)}[\beta_{1}]-\frac{1}{2}N\mathcal{W}^{(3)}[\beta_{1}]^{\T}\beta_{2}\nonumber \\
\qquad & =\beta_{1}N\,K.\label{eq:first_eos_1loop}
\end{align}
The second equation of state \eqref{eq:second_eq_of_state} is
\begin{align}
\frac{\delta\Gamma_{\text{1-loop}}[\beta_{1},\beta_{2}]}{\delta\beta_{2}}= & -\frac{1}{2}N\,\mathcal{W}^{(2)}[\beta_{1}]-\frac{1}{2}\beta_{2}^{-1}\nonumber \\
= & \frac{1}{2}NK.\label{eq:second_eos_1loop}
\end{align}
The derivatives of $\mathcal{W}$ evaluated at $y=\beta_{1}$ by \prettyref{eq:def_W_cal_y}
take the form of the cumulants of $f[z]$ taken with the measure 
\begin{align}
P[z]\propto & e^{S_{0}[z]+\beta_{1}^{\T}f[z]}.\label{eq:measure_z}
\end{align}
Two things are important to note about this measure. First, $\beta_{1}$
has only two non-vanishing components. This means we get $\beta_{1}^{\T}f[z]=-R^{\ast\T}\tilde{x}+\frac{1}{2}\tilde{x}^{\T}Q^{\ast}\tilde{x}$
which is at most quadratic in $z$, as both terms containing $\phi(x)$
vanish. Therefore, the measure \prettyref{eq:measure_z} is Gaussian
which greatly simplifies the calculations. Second, this measure is
not determined by the fluctuation-corrected statistics but the saddle-point
values of the auxiliary fields: $R^{\ast}$ and $Q^{\ast}$. To avoid
confusion, we will use the subscript $\ast$ for cumulants taken with
measure \eqref{eq:measure_z}.

\subsection{Evaluating the 1-loop Equations of State}

We will separate the different contributions to the cumulant by commas
due to the third and fourth entry of $f$ consisting of two parts
with two time arguments. A quick example of this necessity is the
comparison between $\ll f_{4}(s,t)[z]\rr_{\ast}$ and $\ll f_{2}[z](s),f_{2}[z](t)\rr_{\ast}$
because without a separator they look identical: $\ll\phi(s)\phi(t)\rr_{\ast}$
(neglecting prefactors) but this is of course misleading.

With this notation we now close the self-consistency loop by solving
the equations of state for the cumulants. We will start by solving
\eqref{eq:second_eos_1loop} for $\beta_{2}$ which appears linearly,
\begin{align}
\left(\beta_{2}^{-1}\right)_{i,j} & =N\,K_{i,j}+N\,\ll f[z]_{i},f[z]_{j}\rr_{\ast}.\label{eq:2nd_eq_state}
\end{align}
Working under the assumption of a point symmetric activation functions
and under the assumption that $\langle x\rangle=0$, we have $\ll\phi(x)\rr_{\ast}=0$
as well as $\ll\phi^{3}(x)\rr_{\ast}=0$ and $\ll\phi,\tilde{x}\tilde{x}\rr_{\ast}=0$;
the latter is the response of the mean $\langle\phi\rangle$ to a
perturbation of the variance of $x$. Taking into account that any
expectation value solely composed of powers of $\tx$ must vanish,
we see that $\ll f[z]_{i},f[z]_{j}\rr_{\ast}=0$ if $i\in\{1,2\}$
and $j\in\{3,4\}$ or vice versa. Due to the block-diagonal shape
of $K$, $\beta_{2}^{-1}$ is block-diagonal as well. We can therefore
invert these blocks independently. The upper left block of \eqref{eq:2nd_eq_state}
takes the form 
\begin{align*}
(\beta_{2}^{-1})_{11}(t,s) & =N\,\ll\tx(t),\tx(s)\rr_{\ast}=0\\
(\beta_{2}^{-1})_{12}(t,s) & =N\delta(t-s)+N\gb\,\l\tx(t)x(s)\r_{\ast}\l\phi^{\prime}\r_{\ast}\\
(\beta_{2}^{-1})_{21}(t,s) & =N\delta(t-s)+N\gb\,\l\tx(s)x(t)\r_{\ast}\l\phi^{\prime}\r_{\ast}\\
(\beta_{2}^{-1})_{22}(t,s) & =N\gb^{2}\,\ll\phi(t),\phi(s)\rr_{\ast},
\end{align*}
which we can rewrite in momentum-space
\begin{align*}
(\beta_{2}^{-1})_{21}(\omega) & =N-N\gb\frac{\l\phi^{\prime}\r_{\ast}}{1+i\omega},\\
(\beta_{2}^{-1})_{12}(\omega) & =N-N\gb\frac{\l\phi^{\prime}\r_{\ast}}{1-i\omega},\\
(\beta_{2}^{-1})_{22}(\omega) & =N\gb^{2}\ll\phi,\phi\rr_{\ast}(\omega).
\end{align*}
Here we used the results from \prettyref{app:factoring_phi_expectations}
to rewrite $\ll\tilde{x}\phi\rr_{\ast}=\l\phi^{\prime}\r_{\ast}\l\tilde{x}x\r_{\ast}$
and the Fourier representation of the response functions $\l\tilde{x}x\r_{\ast}(\w)=-1/(1+i\w)$,
i.e., the response of a neuron to a $\delta$ perturbation with respect
to the measure \eqref{eq:measure_z}, which has the same form as for
isolated neuron, because the additional term $\beta_{1}^{\T}f(z)$
in the action corresponds to an additional input which does not affect
the response. Finally, we invert this matrix (greatly simplified due
to $(\beta_{2}^{-1})_{11}(t,s)=0$) to find 
\begin{align*}
\beta_{12}(\omega) & =\left((\beta_{2}^{-1})_{21}(\omega)\right)^{-1}\\
 & =N^{-1}\,\frac{1+i\omega}{1-\gb\l\phi^{\prime}\r+i\omega}\\
\beta_{21}(\omega) & =\beta_{12}(-\omega)\\
\beta_{11}(\omega) & =\beta_{12}(\omega)\,(\beta^{-1})_{22}(\omega)\,\beta_{21}(\omega)\\
 & =\frac{1+\omega^{2}}{(1-\gb\l\phi^{\prime}\r)^{2}+\omega^{2}}\frac{\gb^{2}}{N}\ll\phi,\phi\rr_{\ast}(\omega).
\end{align*}
Here we see the first clear sign of the emerging large time constant
in $\beta_{11}(\w)$. When $\bar{g}$ approaches $\l\phi^{\prime}\r^{-1}$
a pole emerges at $\w=0$. This implies that $\beta_{11}(\t)$, the
autocorrelation of the population averaged activity, decays slower
and slower to zero as a function of $t-s$ and thus obtains a large
decay constant. We can also see that $\beta_{22}=0$ as it should
since it is the second cumulant of $\tilde{R}$ which is a response
field. By the same argument $\beta_{44}=\ll\tilde{Q}^{2}\rr$ must
vanish. This implies that one could apply the same method to invert
the lower right block; here we refrain from doing this because our
main interest lies in studying the effect of fluctuations of the population-averaged
activity $R$, which is described by the upper left block.

Next, we solve for the mean via the first equation of state \eqref{eq:first_eos_1loop}
which takes the form 
\begin{align}
(K\beta_{1})_{i} & =-\ll f_{i}[z]\rr_{\ast}-\frac{1}{2}\sum_{l,m\in\underline{4}}\ll f_{i}[z],f_{l}[z],f_{m}[z]\rr_{\ast}\beta_{lm},\label{eq:eq_state_general}
\end{align}
where $\underline{4}=\{1,2,3,4\}$. Note that the multiplication with
$K$, does nothing but switch indices $1\leftrightarrow2$ and $3\leftrightarrow4$.
In principle, \eqref{eq:eq_state_general} determines all mean values
of the population dynamic. We are, however, especially interested
in corrections to $R$ and $Q$, the auxiliary fields used in mean
field. Thus, we consider the cases $i=2$ and $i=4$. The first shows
that the correction on the population activity $R$ caused by its
own fluctuations $\beta_{11}$ is mediated by $\l\langle\phi\tx\tx\rangle\r\propto\l\phi^{\prime\prime}\r$
(for details see \prettyref{app:factoring_phi_expectations}), which
vanishes in the paramagnetic regime. This means that that there is
no influence of fluctuations of $R$ on the transition to the ferromagnetic
state. For the second we need the product of $\beta_{2}$ and the
third cumulant of $f$. This leads to $16$ different combinations
of $l$ and $m$. As discussed above, $\beta_{2}$ has several vanishing
entries: the off-diagonal blocks and the auto-correlations of response
fields, $\beta_{22}$ and $\beta_{44}$. This already reduces the
number of terms from $16$ to $6$. Additionally, the term involving
\begin{align*}
\ll f_{4}[z],f_{3}[z],f_{3}[z]\rr_{\ast} & \propto\ll\phi^{2},\tx^{2},\tx^{2}\rr_{\ast}
\end{align*}
vanishes. This can be shown by methods from \prettyref{app:factoring_phi_expectations},
which work similar to Wick's theorem to express those moments as a
polynomial of second cumulants of $x$ and $\tx$, results in a formula
where every term is at least proportional to $\ll\tx,\tx\rr_{\ast}=0$.

For the fourth component of \eqref{eq:eq_state_general}, this leaves
us with
\begin{align}
Q^{\ast}(s,t)= & (\beta_{1}(s,t))_{3}=(K\beta_{1}(s,t))_{4}\label{eq:Q_star_1_loop}\\
= & g^{2}\ll\phi^{2}(s,t)\rr_{\ast}\label{eq:mf_contrib}\\
 & +\frac{1}{2}g^{2}\int_{u,v}\ll\phi^{2}(s,t),\tx(u),\tx(v)\rr_{\ast}\beta_{11}(u,v)\label{eq:beta11_contrib}\\
 & +g^{2}\gb\int_{u,v}\ll\phi^{2}(s,t),\tx(u),\phi(v)\rr_{\ast}\beta_{12}(u,v)\label{eq:beta12_contrib}\\
 & -\frac{g^{4}}{2}\int_{u_{1,2},v_{1,2}}\ll\phi^{2}(s,t),\tx^{2}(u_{1},u_{2}),\phi^{2}(v_{1},v_{2})\rr_{\ast}\nonumber \\
 & \phantom{-\frac{g^{4}}{2}\int_{u_{1,2},v_{1,2}}}\beta_{34}(u_{1},u_{2},v_{1},v_{2}).\label{eq:beta34_contrib}
\end{align}
This equation lends itself nicely to interpretation using the intuitive
picture of a mean-field neuron embedded in a `bath' of activity due
to the network (akin to the cavity method \citep{Mezard87}). The
first contribution \eqref{eq:mf_contrib} is identical to the mean-field
approximation. The next contribution \eqref{eq:beta11_contrib} contains
$\beta_{11}$, the autocorrelation of the population averaged activity
$R$. This term can be interpreted as the effect of fluctuations of
$R$ measured by $\beta_{11}$ contributing to the variance of the
input of the representative mean-field neuron. Term \eqref{eq:beta12_contrib}
shows how a fluctuation of the neuronal activity $\phi(v)$ is echoed
in the network and transmitted back by the response function $\beta_{12}$
of the bath, affecting the mean input by coupling to $\tx(u)$ which,
in turn, modifies the variance of the mean-field neuron's input by
changing the second moment $\langle\phi^{2}(s,t)\rangle$. Similarly
\eqref{eq:beta34_contrib} shows an echo effect: A fluctuation of
$\phi^{2}(v_{1},v_{2})$ propagates through the bath with the response
$\beta_{34}(u_{1},u_{2},v_{1},v_{2})$ to time points $u_{1},u_{2}$
and causes a change of the variance in the input of the mean-field
neuron by coupling to $\tx^{2}(u_{1},u_{2})$, which in turn affects
$\langle\phi^{2}(s,t)\rangle$.

\section{Results}

For this section we consider the regime $g<1$ and set $\phi(x)=\erf(\ensuremath{\sqrt{\pi}}x/2)$
which makes all involved expectation values of $\phi$ and its derivatives
as they appear in \prettyref{app:factoring_phi_expectations} solvable
analytically \citep{Owen80_389,vanMeegen21_043077} while staying
close to the popular choice of a hyperbolic tangent. Furthermore,
we only consider the corrections \eqref{eq:beta11_contrib} due to
$\beta_{11}$ which empirically dominates the other contributions
(for an explicit expression for $Q^{\ast}$ including the contributions
due to $\beta_{12}$ in linear networks see \prettyref{app:linear_analytical}).

\begin{figure}
\includegraphics[width=1\columnwidth]{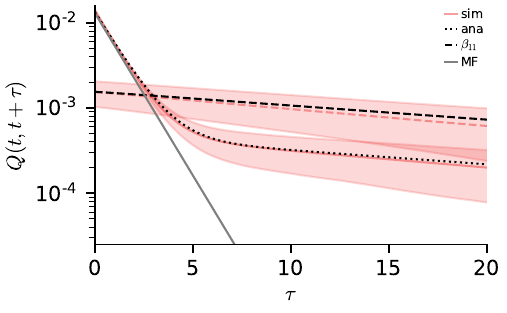}

\caption{Time lagged population-averaged autocorrelation $Q(t,t+\tau)$ \eqref{eq:def_Q}
simulated (red) and self consistent solution \prettyref{eq:Q_star_1_loop}
(black) together with autocorrelation $\protect\ll R(t+\tau)R(t)\rangle\rangle=\beta_{11}(\tau)$
of population-averaged activity $R$ \eqref{eq:def_R} (dashed, self
consistent in black, empirical in red) plotted logarithmically for
$\bar{g}=1.0$. Other parameters as in \prettyref{fig:R}.}
\label{fig:autocorrelation_analytical}
\end{figure}

\prettyref{fig:autocorrelation_analytical} shows the autocorrelation
for a network close to the phase transition. In the simulation results
we observe the critical slowing down already visible in \prettyref{fig:R}c,
which our self-consistent theory describes quite well. Above all,
we see the emerging time constant corresponding to the decay of the
network activities' autocorrelation $\beta_{11}(t-s)=\l\l R(t)\,R(s)\r\r.$
Also the autocorrelation function features two different time scales:
The fast time-scale dominates the initial decay for time lags close
to zero; this part is identical to the mean-field result neglecting
fluctuations. The second time scale dominates the behavior of the
autocorrelation function at large time lags. Its is caused by the
fluctuations of $R$ as quantified by $\beta_{11}$.

\begin{figure}
\includegraphics[width=1\columnwidth]{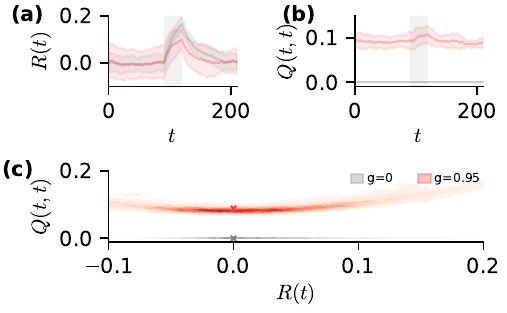}\caption{(\textbf{a}) Transient of $R$ in response to a stimulation provided
as common input of $0.01$ to each neuron (additive constant on right
hand side of \prettyref{eq:diffeq_motion}) within the time span indicated
by the shaded region; $\bar{g}=1$ and different values of $g$ (colors
given in legend) (\textbf{b}) Transient of $Q$ under same conditions
as in a. (\textbf{c}) 2D histogram of $Q$ over $R$ with crosses
at the zero time lag predicted as $Q^{\ast}(t,t)$ from theory \eqref{eq:Q_star_1_loop}.
Other parameters as in \prettyref{fig:R}. \label{fig:QR_const}}
\end{figure}
\prettyref{fig:QR_const} shows how a network's response to constant
input changes close to the transition for different values of $g$.
The population activity of a network with no variance in its connection
($g=0$) behaves like a capacitor. For $g>0$, the increase of the
population activity due to the transient input is suppressed compared
to $g=0$. This highlights that close to the transition to the chaotic
regime, a rise in the population-averaged activity $R$ is counteracted
by the increase of the variance measured by $Q$; formally this can
be seen from the effective slope of the noise-averaged activation
function (cf. \eqref{eq:SC_erf_mu}) to decrease with increasing $Q$,
which in turn reduces the positive feedback that controls the dynamics
of $R$ by \eqref{eq:eta_mean}. This stronger variability and the
coupling of $R$ and $Q$ can be seen in \prettyref{fig:QR_const}c
in the higher curvature for larger $g$. Our theory captures the resulting
slightly elevated average of $Q$, as can be seen by the analytical
crosses indicating $Q(\tau=0)$ lying slightly above the parabolas'
low points.

\begin{figure}
\includegraphics[width=1\columnwidth]{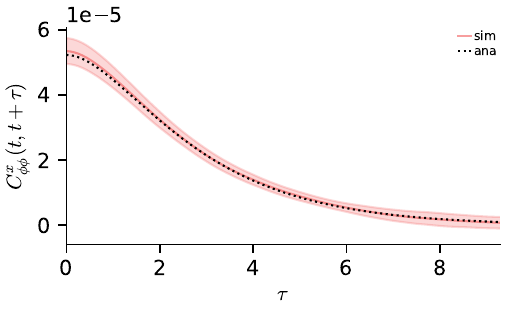}\caption{Population averaged cross correlation $C_{\phi\phi}^{x}(\tau)$ \prettyref{eq:cross_corr}
over time lag given by \prettyref{eq:cross_corr} (black) compared
to simulation (red) for $\bar{g}=0.5$. Other parameters as in \prettyref{fig:R}.\label{fig:cross_correlation}}
\end{figure}
Direct access to $Q$ and the fluctuations of $R$ also allows us
to conveniently calculate the pairwise averaged cross-correlation
of the output 
\begin{align}
C_{\phi\phi}^{x}(t-s):=\frac{1}{N^{2}}\sum_{i\neq j}\phi_{i}(s)\phi_{j}(t)= & \frac{\beta_{11}(s,t)}{\bar{g}^{2}}-\frac{Q(s,t)}{Ng^{2}}\label{eq:cross_corr}
\end{align}
as can be seen in \prettyref{fig:cross_correlation}. \prettyref{eq:cross_corr}
highlights the large time constant present in the cross correlation
induced by the network level correlation $\beta_{11}$ which was also
shown by \citet{Clark23_118401} using cavity methods.

\section{Discussion}

In this paper we investigated the critical behavior close to the structured
(ferromagnetic) regime of the Sompolinsky-Crisanti-Sommers model
with non-zero mean connectivity and noise. After first reproducing
the phase diagram using (dynamical) mean-field theory \citep{Mastrogiuseppe17_e1005498},
we derive a self-consistent set of equations to one loop order, systematically
taking corrections of order $1/N$ into account. Our theory explains
the emergence of long time scales in the decay of the population averaged
autocorrelation function $Q$, which we show to be caused by fluctuations
of the population-averaged population activity $R$. The theory furthermore
links these network level effects to pairwise correlations on the
single neuron scale. We thus successfully bridge between the emerging
large timescales of the autocorrelation on the single neuron scale
and finite size effects on the network level. Lastly, our analytical
results explain how fluctuations of the population-averaged activity
lead to a higher population averaged autocorrelation, showing a correlation
in the two auxiliary fields that span the phase space of recurrent
networks and are conventionally studied in mean-field theory.

With regard to the study of criticality in neuronal networks, we have
provided a model that features two critical transitions. First, the
transition between the regular regime and the chaotic phase, which
is predominantly controlled by the amount of disorder in the connectivity
quantified by $g$ and, in the absence of driving noise, indicated
by the order parameter $Q$. This transition has been studied extensively
in many previous works \citep{Sompolinsky88_259,Marti18_062314,Schuecker18_041029}.
Our analysis here focuses on the ``ferromagnetic'' transition mainly
controlled by the parameter $\bar{g}$, for which $R$ plays the role
of an order parameter. Our theory explicitly demonstrates critical
slowing down of the dynamics at the point of the continuous phase
transition and allows the computation of the time scale. The theory,
moreover, exposes that the two transitions cannot be studied in isolation,
because we find a tight interplay of the two order parameters: fluctuations
of $R$ directly affect the order parameter $Q$, in particular the
latter inherits the slow temporal decay from the critical fluctuations
of the former. Also vice versa, the response of $R$ is found to be
multi-phased, which appears to be caused by the back influence of
$Q$ on $R$.

On the side of network theory, the proposed method of second-order
Legendre transform to obtain a renormalized theory in the form of
a set of self-consistency equations for the first and second-order
statistics of the population activity may be useful to study other
network properties. For example, within the framework of Bayesian
inference \citep{Williams98_1203,Lee18}, one cornerstone of contemporary
theory of deep neuronal networks \citep{ZavatoneVeth21_NeurIPS_I,ZavatoneVeth21_NeurIPS_II,seroussi23_908},
the presented theory may be useful to compute the network prior. An
interesting feature in this regard is that the neurons in our renormalized
theory do not decouple, in contrast to the case of the large $N$-limit
for deep and recurrent networks with centered prior distributions
on the weights \citep{Segadlo22_103401}. We hope that the presented
framework will be useful to understand the functional consequences
of this finding and that it will open the door to studying the finite-size
properties of recurrent stochastic networks in continuous time in
general.
\begin{acknowledgments}
We are grateful for helpful discussions with Andrea Crisanti in the
early stages of this project and Tobias Kühn for valuable feedback
on the manuscript. This project has received funding from the European
Union\textquoteright s Horizon 2020 Framework Programme for Research
and Innovation under Specific Grant Agreement No. 945539 (Human Brain
Project SGA3); the Helmholtz Association: Young investigator's grant
VH-NG-1028; the German Federal Ministry for Education and Research
(BMBF Grant 01IS19077A to Jülich); Open access publication funded
by the Deutsche Forschungsgemeinschaft (DFG, German Research Foundation)
-- 491111487.  MD received funding as Vernetzungsdoktorand: ``Dynamic
characteristics of reservoir computing''
\end{acknowledgments}

\bibliographystyle{apsrev4-2_prx}
%

\section*{Appendix}

\subsection{Critical Coupling Strength in Mean-Field Theory\label{app:critical_coupling}}

We choose $\phi(x)=\mathrm{erf}(\sqrt{\pi}x/2)$, where the scaling
ensures $\phi^{\prime}(0)=1$, for which the expectations on the r.h.s.~of
\eqref{eq:eta_mean} and \eqref{eq:eta_corr} are solvable analytically
\citep[III.A.3]{vanMeegen21_043077}. In the stationary state, they
are

\begin{align}
\mu_{\phi} & =\phi\Big(\frac{\mu_{x}}{\sqrt{1+\tfrac{\pi}{2}\sigma_{x}^{2}}}\Big),\label{eq:SC_erf_mu}\\
C_{\phi\phi}(\tau) & =1-8T\Big(\frac{\sqrt{\tfrac{\pi}{2}}\mu_{x}}{\sqrt{1+\tfrac{\pi}{2}\sigma_{x}^{2}}},\frac{\sqrt{1+\tfrac{\pi}{2}\sigma_{x}^{2}(1-\rho_{x}(\tau))}}{\sqrt{1+\tfrac{\pi}{2}\sigma_{x}^{2}(1+\rho_{x}(\tau))}}\Big),\label{eq:SC_erf_C}
\end{align}
with $\sigma_{x}^{2}=C_{x}(0)$, $\rho_{x}(\tau)=C_{x}(\tau)/\sigma_{x}^{2}$,
and Owen's T function $T(h,a)=\frac{1}{2\pi}\int_{0}^{a}dx\,(1+x^{2})^{-1}e^{-\frac{1}{2}h^{2}(1+x^{2})}$.

Inserting $\mu_{x}=\bar{g}\mu_{\phi}$ into \prettyref{eq:SC_erf_mu},
we obtain
\begin{align}
\mu_{x} & =\bar{g}\,\phi\Big(\frac{\mu_{x}}{\sqrt{1+\tfrac{\pi}{2}\sigma_{x}^{2}}}\Big).\label{eq:SC_erf_mu_x}
\end{align}
Since $\phi(x)$ is sigmoidal and symmetric, \prettyref{eq:SC_erf_mu_x}
has either one or three solutions---the latter corresponds to the
state with a broken symmetry.

We approach the transition from the symmetric domain with $\mu_{\phi}=0$.
The deciding criterion to make multiple solutions possible is a unit
slope at zero,
\begin{align}
\frac{\bar{g}_{c}}{\sqrt{1+\tfrac{\pi}{2}\sigma_{x}^{2}}} & =1.\label{eq:crit_criterion}
\end{align}
For $\mu_{x}=0$, \prettyref{eq:SC_erf_C} simplifies to $C_{\phi\phi}(\tau)=\frac{2}{\pi}\,\arcsin y(\tau)$
where $y(\tau)=(1+\tfrac{\pi}{2}\sigma_{x}^{2})^{-1}\tfrac{\pi}{2}\sigma_{x}^{2}\rho_{x}(\tau)$.
This leads to the differential equation $\ddot{y}=-\partial_{y}V(y,y_{0})$
with \citep{vanMeegen21_158302}
\begin{equation}
V(y,y_{0})=-\frac{1}{2}y^{2}+g^{2}(1-y_{0})\left(\sqrt{1-y^{2}}+y\arcsin\left(y\right)-1\right).
\end{equation}
Energy conservation determines the initial condition $y_{0}=(1+\tfrac{\pi}{2}\sigma_{x}^{2})^{-1}\tfrac{\pi}{2}\sigma_{x}^{2}$
and leads to $D=\frac{2}{\pi}(1-y_{0})^{-1}\sqrt{-2V(y_{0},y_{0})}.$
Using the stability criterion \prettyref{eq:crit_criterion} yields
$y_{0}=1-\bar{g}_{c}^{-2}$ and thus
\begin{equation}
D=\frac{2}{\pi}\bar{g}_{c}^{2}\sqrt{-2V(1-\bar{g}_{c}^{-2},1-\bar{g}_{c}^{-2})},\label{eq:D_crit}
\end{equation}
where the dependence on $g$ is in $V(y,y_{0})$. We obtained $D=D(\bar{g}_{c},g^{2})$
which can be solved numerically for $\bar{g}_{c}=\bar{g}_{c}(g,D)$.

\subsection{Double Legendre Transformation in Cumulants\label{app:legendre_in_cumulants}}

\label{appendix:double_legendre_transform}To derive the Legendre
transformation in terms of cumulants, we restrict ourselves to a second-order
transformation for brevity. We start by studying the properties of
the effective action resulting from a first order Legendre transformation
to establish a ground truth. Then we will define the transformation
in cumulants and compare its properties.

For the sake of keeping calculations concise we dress the second source
term with a factor $\frac{1}{2}$ in the cumulant generating functional
$W[j,k]=\ln\int_{y}\e\left(S[y]+j^{\T}y+\frac{1}{2}y^{\T}ky\right)$
which gives us the moments
\begin{align*}
\alpha_{1} & =\l y\r=\partial_{j}W,\\
\alpha_{2} & =\frac{1}{2}\l y^{2}\r=\partial_{k}W.
\end{align*}
From here we continue to obtain the effective action
\begin{align*}
\Gamma_{m}[\alpha_{1},\alpha_{2}] & =\text{extr}_{j,k}\,j^{\T}\alpha_{1}+k^{\T}\alpha_{2}-W[j,k],
\end{align*}
which we decorated with a subscript $m$ for ``moment'' to differentiate
it from the effective action constructed via cumulants. This yields
the corresponding equations of state
\begin{align*}
\frac{\partial}{\partial\alpha_{1}}\Gamma_{m}[\alpha_{1},\alpha_{2}] & =j,\\
\frac{\partial}{\partial\alpha_{2}}\Gamma_{m}[\alpha_{1},\alpha_{2}] & =k.
\end{align*}
Our goal now is to reformulate these expressions in terms of cumulants
instead of moments. To this end we consider the definition of cumulants:
\begin{align*}
\beta_{1} & :=\partial_{j}W=\l y\r=\alpha_{1}\\
\beta_{2} & :=\partial_{j}^{2}W=\l y^{2}\r-\l y\r^{2}=2\alpha_{2}-\alpha_{1}^{2},
\end{align*}
which can be inverted to yield
\begin{align*}
\alpha_{1}(\beta_{1}) & =\beta_{1},\\
\alpha_{2}(\beta_{1},\beta_{2}) & =\frac{1}{2}(\beta_{2}+\beta_{1}^{2}).
\end{align*}
This gives us a straightforward way of defining the Legendre transformation
as a function of cumulants in terms of the standard Legendre transformation
as a function of moments as 
\begin{align}
\Gamma_{c}[\beta_{1},\beta_{2}] & :=\Gamma_{m}\left[\beta_{1},\,\frac{1}{2}\left(\beta_{2}+\beta_{1}^{2}\right)\right]\label{eq:Gamma_c_def}
\end{align}
and its equations of state
\begin{align*}
\partial_{\beta_{1}}\Gamma_{c}[\beta_{1},\beta_{2}]= & \partial_{\beta_{1}}\Gamma_{m}[\beta_{1},\frac{1}{2}(\beta_{2}+\beta_{1}^{2})]\\
= & \underbrace{\partial_{1}\Gamma_{m}[\beta_{1},\frac{1}{2}(\beta_{2}+\beta_{1}^{2})]}_{j}\\
 & +\underbrace{\partial_{2}\Gamma_{m}[\beta_{1},\frac{1}{2}(\beta_{2}+\beta_{1}^{2})]}_{k}\beta_{1}\\
= & j+\beta_{1}^{\T}k,\\
\partial_{\beta_{2}}\Gamma_{c}[\beta_{1},\beta_{2}]= & \partial_{\beta_{2}}\Gamma_{m}[\beta_{1},\frac{1}{2}(\beta_{2}+\beta_{1}^{2})]\\
= & \frac{1}{2}\underbrace{\partial_{2}\Gamma_{m}[\beta_{1},\frac{1}{2}(\beta_{2}+\beta_{1}^{2})]}_{k}\\
= & \frac{1}{2}k,
\end{align*}
in terms of the cumulants. Here $\partial_{1}$ and $\partial_{2}$
describe the derivative with respect to the first and second variable.
To make the cumulant dependency more explicit we rewrite \eqref{eq:Gamma_c_def}:
\begin{align*}
\Gamma_{c} & [\beta_{1},\beta_{2}]\\
 & =\text{extr}_{j,k}\,-\ln\int_{y}e^{S[y]+j^{\T}y+\frac{1}{2}y^{\T}ky-j^{\T}\beta_{1}-k^{\T}\frac{1}{2}(\beta_{2}+\beta_{1}^{2})}\\
 & =\text{extr}_{j,k}\,-\ln\int_{y}e^{S[y]+(j+k^{\T}\beta_{1})^{\T}(y-\beta_{1})+\frac{k^{\T}}{2}\left((y-\beta_{1})^{2}-\beta_{2}\right)}.
\end{align*}
By introducing $\hat{j}:=j+k^{\T}\beta_{1}$ we keep the form of the
stationarity equation with $\tfrac{\partial}{\partial\hat{j}}$ applied
to the right hand side yielding zero. Both pairs of stationarity conditions
$\{\partial/\partial j$, $\partial/\partial k\}$ and $\{\partial/\partial\hat{j},\partial/\partial k\}$
imply the same pair of constraints because the only additional term
produced by $\partial/\partial k$ acting on $(k^{\T}\beta_{1})^{\T}(y-\beta_{1})$
is proportional to $\langle y-\beta_{1}\rangle$ which, as a result
of the constraint enforced by $\partial/\partial j$, vanishes. We
can thus write the effective action for cumulants as 
\begin{align}
\Gamma_{c}[\beta_{1},\beta_{2}] & =\text{extr}_{\hat{j},k}\,-\ln\int_{y}e^{S[y]+\hat{j}^{\T}(y-\beta_{1})+\frac{k^{\T}}{2}\left((y-\beta_{1})^{2}-\beta_{2}\right)},\label{eq:Gamma_c}
\end{align}
with 
\begin{align*}
\partial_{\hat{j}}\Gamma_{c}[\beta_{1},\beta_{2}] & =0,\\
\partial_{k}\Gamma_{c}[\beta_{1},\beta_{2}] & =0,\\
\partial_{\beta_{1}}\Gamma_{c}[\beta_{1},\beta_{2}] & =j+\beta_{1}^{\T}k,\\
\partial_{\beta_{2}}\Gamma_{c}[\beta_{1},\beta_{2}] & =\frac{1}{2}k.
\end{align*}
We have thus found a way to describe a system in terms of cumulants.

\subsection{Factoring $\phi$-Expectation Values\label{app:factoring_phi_expectations}}

In this chapter we show how to calculate $\l\prod_{i}^{n}\tx(s_{i})\prod_{j}^{m}\phi(x(t_{j}))\r$
under a Gaussian measure, which we need to solve our self consistent
equations. We will do so inspired by the derivations of Price Theorem
\citep{Papoulis91}. This means we express $\phi$ via their Fourier
representation and pull everything except the exponential necessary
for this transformation outside of the expectation value. Doing so
comes at the cost of having to calculate some derivatives, but this
enables us to perform the integration for the expectation values and
in the end clean up by calculating said derivatives.

The first step is the Fourier representation and insertion of a source
term for $\tx$. We then replace $\tx$ by derivatives with respect
to this source, allowing us to rewrite $\l\prod_{i}^{n}\tx(s_{i})\prod_{j}^{m}\phi(x(t_{j}))\r$
as:
\begin{align}
 & \prod_{j}\left\{ \int dk_{j}\hp(k_{j})\right\} \left\langle \prod_{i}\left\{ \tx(s_{i})\right\} \exp(\i\sum_{j}\kj x(t^{\prime})dt^{\prime})\right\rangle \nonumber \\
= & \prod_{j}\left\{ \int dk_{j}\hp(k_{j})\right\} \Big[\prod_{i}\left\{ -\i\delta_{\tilde{k}(s_{i})}\right\} \label{eq:phi_exp_factor}\\
 & \quad\underbrace{\left\langle \exp(\i\sum_{j}\kj x(t^{\prime})dt^{\prime}+\i\tilde{k}\tx)\right\rangle }_{(\star)}\Big]_{\tilde{k}=0}\nonumber 
\end{align}
There we used the abbreviations $\kj:=\int2\pi k_{j}\delta(t_{j}-t^{\prime})$.
From here we focus on the expectation value denoted by $\star$ which
we write out as an integration over $x$ and $\tx$ w.r.t.~their
distribution function. We assume the fields to be distributed according
to a Gaussian with covariance 
\begin{align*}
\Delta & =\left(\begin{array}{cc}
\Delta_{xx} & \Delta_{x\tilde{x}}\\
\Delta_{x\tilde{x}}^{\ast} & 0
\end{array}\right).
\end{align*}
Here it is important to note that $\tx$ is a response-field and therefore
has a vanishing autocorrelation, allowing us to set $\Delta_{\tilde{x}\tilde{x}}=0$.
The resulting integrand is an exponential of a quadratic polynomial,
which we solve using a Hubbard Stratonovich Transformation: 
\begin{align*}
(\star)= & \int dx\,d\tx\exp(-\frac{1}{2}\left(\begin{array}{c}
x\\
\tx
\end{array}\right)^{\T}\Delta^{-1}\left(\begin{array}{c}
x\\
\tx
\end{array}\right)\\
 & \phantom{\int dx\,d\tx\exp(}+\i x\sum_{j}\kj+\i\tilde{k}\tx)\\
 & =\exp(-\frac{1}{2}\left(\begin{array}{c}
\sum\kj\\
\tilde{k}
\end{array}\right)^{\T}\Delta\left(\begin{array}{c}
\sum\kj\\
\tilde{k}
\end{array}\right))\\
 & =:\exp(D(\tilde{k})).
\end{align*}
Plugging this back into \ref{eq:phi_exp_factor}, we see that the
derivatives we need to calculate are $\prod_{i}\left\{ \delta_{\tilde{k}(s_{i})}\right\} \exp(D(\tilde{k}))$.
As $D(\tilde{k})$ is just a quadratic equation, we can easily calculate
its first two derivatives: 
\begin{align*}
\delta_{\tilde{k}(s_{i})}D(\tilde{k}) & =D^{\prime}(\tilde{k})(s_{i})\\
 & =-\left(\begin{array}{c}
0\\
1
\end{array}\right)^{\T}\Delta\left(\begin{array}{c}
\sum_{j}\kj\\
\tilde{k}
\end{array}\right)\\
 & =-\int_{t^{\prime}}\Delta_{x\tx}(s_{i},t^{\prime})\sum_{j}\kj\\
 & =-\sum_{j}\Delta_{x\tx}(s_{i},t_{j})k_{j}\\
\Rightarrow\qquad D^{\prime\prime}(\tilde{k}) & =0.
\end{align*}
Thus it can be shown by induction: 
\begin{align*}
\prod_{i}\left\{ \delta_{\tilde{k}(t_{i})}\right\} \exp(D(\tilde{k}))= & \exp(D(\tilde{k}))\,\prod_{i}D^{\prime}(\tilde{k})(s_{i})
\end{align*}
with base case
\begin{align*}
\exp(D(0)) & =\exp(-\frac{1}{2}\sum\kj\Delta_{11}\sum\kj)\\
 & =\int dx\,d\tx\exp(-\frac{1}{2}\left(\begin{array}{c}
x\\
\tx
\end{array}\right)^{\T}\Delta^{-1}\left(\begin{array}{c}
x\\
\tx
\end{array}\right)\\
 & \phantom{=\int dx\,d\tx\exp(}+\i x\sum_{j}\kj)\\
 & =\left\langle \exp(\i x\sum_{j}\kj)\right\rangle \\
 & =\left\langle \prod_{j}\exp(\i x\kj)\right\rangle \\
 & =\left\langle \prod_{j}\exp(\i\,k_{j}\,x(t_{j}))\right\rangle .
\end{align*}
This way we get: 
\begin{align*}
\l\prod_{i}^{n}\tx(s_{i})\prod_{j}^{m}\phi(x(t_{j}))\r= & \l\prod_{j}\left\{ \int dk_{j}\hp(k_{j})\exp(\i\,k_{j}\,x(t_{j}))\right\} \r\\
 & \cdot\underbrace{\prod_{i}^{n}\sum_{j}^{m}\Delta_{x\tx}(s_{i},t_{j})\i k_{j}}_{\left(\star\star\right)}.
\end{align*}
To simplify this further, we want to perform the $k_{j}$-integrals
but for this need to rewrite the term denoted by $(\star\star)$.
One way of doing this would be: 
\begin{align*}
(\star\star)= & \left(\sum_{j_{1}}^{m}\Delta_{x\tx}(s_{i},t_{j_{1}})\i k_{j_{1}}\right)\dots\left(\sum_{j_{n}}^{m}\Delta_{x\tx}(s_{i},t_{j_{n}})\i k_{j_{n}}\right)\\
= & \sum_{\bm{j}}\prod_{i}^{n}\Delta_{x\tx}(s_{i},t_{j_{i}})\i k_{j_{i}},
\end{align*}
Even though this makes it very cumbersome to perform the integral,
as the multi-indexed $k$ can be part of any integral, this form lets
us see how the final result must look like: each term of the sum has
to have one factor $\prod_{j}k_{j}^{\alpha_{j}}$ with $\sum_{j}\alpha_{j}=n$.
This is then multiplied by a sum of products of $\Delta_{x\tx}(s_{i},t_{j})$,
where each $j$ appears as often in each product as the value of $\alpha_{j}$
and the $i$ takes all values from $1$ to $n$. The sum then goes
over all different combinations of $i$ and $j$. So we introduce
\begin{align*}
j(\alpha)=(\underbrace{1,\dots,1}_{\alpha_{1}\text{times}},\underbrace{2\dots,2}_{\alpha_{2}\text{times}},\underbrace{3,\dots,3}_{\alpha_{3}\text{times}},\dots),\\
j(\alpha,i)=(j(\alpha))_{i},
\end{align*}
and the permutation group $S$ which does not switch the $j$'s which
are equal. Doing so allows us to finally rewrite $\l\prod_{i}^{n}\tx(s_{i})\prod_{j}^{m}\phi(x(t_{j}))\r$:
\begin{align*}
 & \l\prod_{j}\left\{ \int dk_{j}\hp(k_{j})\exp(\i\,k_{j}\,x(t_{j}))\right\} \r\\
\cdot & \sum_{\alpha,\,\sum_{j}\alpha_{j}=n}\prod_{j}\left\{ \left(\i k_{j}\right)^{\alpha_{j}}\right\} \sum_{\sigma\in S}\prod_{i}\Delta_{x\tx}(s_{i},t_{\sigma(j(\alpha,i))})\\
= & \sum_{\alpha,\,\sum_{j}\alpha_{j}=n}\l\prod_{j}\left\{ \int dk_{j}\left(\i k_{j}\right)^{\alpha_{j}}\hp(k_{j})\exp(\i\,k_{j}\,x(t_{j}))\right\} \r\\
\cdot & \sum_{\sigma\in S}\prod_{i}\Delta_{x\tx}(s_{i},t_{\sigma(j(\alpha,i))})\\
= & \sum_{\alpha,\,\sum_{j}\alpha_{j}=n}\l\prod_{j}\left\{ \phi^{(\alpha_{j})}(x(t_{j}))\right\} \r\sum_{\sigma\in S}\prod_{i}\Delta_{x\tx}(s_{i},t_{\sigma(j(\alpha,i))}).
\end{align*}
For the relevant cases this results in 
\begin{align*}
\ll\phi(t)\phi(s),\tx(u),\tx(v)\rr & =\l\phi^{2}(s,t)\tx(u)\tx(v)\r\\
 & =\l\phi^{\prime\prime}(s)\phi(t)\r\l\tx(u)x(s)\r\l\tx(v)x(s)\r\\
 & +\l\phi^{\prime\prime}(t)\phi(s)\r\l\tx(u)x(t)\r\l\tx(v)x(t)\r\\
 & +\l\phi^{\prime}(s)\phi^{\prime}(t)\r(\l\tx(u)x(t)\r\l\tx(v)x(s)\rangle\\
 & \qquad+\l\tx(u)x(s)\r,\l\tx(v)x(t)\r)\\
\ll\phi(t)\phi(s),\tx(u),\phi(v)\rr & =\l\phi(s)\phi(t)\tx(u)\phi(v)\r\\
 & -\l\phi(s)\phi(t)\r\l\tx(u)\phi(v)\r\\
 & =\l\phi^{\prime}(s)\phi(t)\phi(v)\r\l\tx(u)x(s)\r\\
 & +\l\phi^{\prime}(t)\phi(s)\phi(v)\r\l\tx(u)x(t)\r\\
 & +\l\phi^{\prime}(v)\phi(s)\phi(t)\r\langle\tilde{x}(u)x(v)\rangle\langle\phi^{\prime}\r\\
 & -\l\phi(s)\phi(t)\r\l\tx(u)\phi(v)\r\\
\ll\phi(t)\phi(s),\phi(u),\tx(v)\rr & =\l\phi(s)\phi(t)\phi(u)\tx(v)\r\\
 & -\l\phi(s)\phi(t)\r\l\tx(v)\phi(u)\r\\
 & =\l\phi^{\prime}(s)\phi(t)\phi(u)\r\l\tx(v)x(s)\r\\
 & +\l\phi^{\prime}(t)\phi(s)\phi(u)\r\l\tx(v)x(t)\r\\
 & +\l\phi^{\prime}(u)\phi(s)\phi(t)\r\l\tx(v)x(u)\r\\
 & -\l\phi(s)\phi(t)\r\langle\tilde{x}(v)x(u)\rangle\langle\phi^{\prime}\r.
\end{align*}

\subsection{Analytical Results for Linear Activation Function\label{app:linear_analytical}}

Here we consider the case $\phi(x)=x$ and derive an explicit solution
for the first equation of state (\ref{eq:Q_star_1_loop}):
\begin{align}
Q^{\ast}(s,t)= & g^{2}\ll x^{2}\rr_{\ast}(s,t)\nonumber \\
 & +\frac{1}{2}g^{2}\int_{u,v}\ll x^{2}(s,t),\tx(u),\tx(v)\rr_{\ast}\beta_{11}(u,v)\nonumber \\
 & +g^{2}\gb\int_{u,v}\ll x^{2}(s,t),\tx(u),x(v)\rr_{\ast}\beta_{12}(u,v).\label{eq:Q_linear}
\end{align}
For this we need values for the second cumulant of the auxiliary fields
and particular third cumulants of the single neuron dynamic. To get
the former we use the first equation of state (\ref{eq:Q_star_1_loop})
in Fourier domain 
\begin{align}
\beta_{12}(\w) & =\frac{1}{N}\frac{1+i\w}{1-\gb+i\w},\label{eq:beta21_linear}\\
\beta_{11}(\w) & =N\beta_{12}(\w)\l xx\r_{1}(\w)\beta_{21}(\w)\label{eq:beta11_linear}\\
 & =\frac{1}{N}\frac{Q(\w)+D}{(1-\gb)^{2}+\w^{2}},\nonumber 
\end{align}
where we used $\l xx\r_{1}(\w)=\frac{Q(\w)+D}{1+\w^{2}}$ in the last
line. The latter we can break down into second moments via the methods
derived in \ref{app:factoring_phi_expectations}
\begin{align*}
\ll x(t)x(s),\tilde{x}(u),\tilde{x}(v)\rr_{\ast} & =\l\tx(u)x(t)\r\l\tx(v)x(s)\rangle\\
 & +\l\tx(u)x(s)\r,\l\tx(v)x(t)\r\\
\ll x(t)x(s),\tx(u),x(v)\rr_{\ast} & =\l x(t)x(v)\r\l\tx(u)x(s)\r\\
 & +\l x(s)x(v)\r\l\tx(u)x(t)\r.
\end{align*}
As we will solve the integrals in Fourier domain, it is easiest to
name them:
\begin{align*}
I_{11}(s-t) & =\frac{g^{2}}{2}\int du\,dv\,\ll x(t)x(s),\tilde{x}(u),\tilde{x}(v)\rr_{\ast}\beta_{11}(u-v)\\
 & =g^{2}\int du\,dv\,\l\tx(u)x(s)\r_{1}\l\tx(v)x(t)\r_{1}\beta_{11}(u-v).
\end{align*}
This leaves us with a double convolution which turns into a product
in Fourier domain: 
\begin{align*}
g^{-2}I_{11}(\tau) & =\int\frac{d\w}{2\pi}\l\tx x\r_{1}(\w)\l\tx x\r_{1}(-\w)\beta_{11}(-\w)\exp(i\w(\tau))\\
 & =\int\frac{d\w}{2\pi}\frac{\beta_{11}(\w)}{1+\w^{2}}\exp(i\w(\tau))\\
 & =\int\frac{d\w}{2\pi}\frac{1}{N}\frac{Q(\w)+D}{(1-\gb)^{2}+\w^{2}}\frac{1}{1+\w^{2}}\exp(i\w(\tau))\\
 & =\int\frac{d\w}{2\pi}I_{11}(\w)\exp(i\w(\tau)),
\end{align*}
whereby we used again the evenness of $\beta_{11}$ and the shape
of the activity's response function in Fourier domain $\l\tilde{x}x\r(\omega)=-(1+i\omega)^{-1}$.

We can deal with the second integral equivalently:
\begin{align*}
\frac{I_{12}(s-t)}{g^{2}\gb} & =\int_{u,v}\ll x^{2}(s,t),\tx(u),x(v)\rr_{\ast}\beta_{12}(u,v)\\
 & =\int_{u,v}\left(\l x(t)x(v)\r\l\tx(u)x(s)\r\right.\\
 & \left.+\l x(s)x(v)\r\l\tx(u)x(t)\r\right)\beta_{12}(u,v)\\
 & =\int_{u,v}\left(\l xx\r(v-t)\l\tx x\r(s-u)\right.\\
 & \left.+\l xx\r(v-s)\l\tx x\r(t-u)\right)\beta_{12}(u,v)\\
 & =\int\frac{d\omega}{2\pi}\exp(i\omega(s-t))\\
 & \cdot\frac{1}{N}\frac{1-\bar{g}+\omega^{2}(1+\bar{g})}{(1-\bar{g})^{2}+\omega^{2}}\frac{-2}{1+\omega^{2}}\l xx\r_{1}(\omega)
\end{align*}
and thus get: 
\begin{align*}
I_{11}(\w) & =\frac{g^{2}}{N}\frac{Q(\w)+D}{(1-\gb)^{2}+\w^{2}}\frac{1}{1+\w^{2}}
\end{align*}
\begin{align*}
I_{12}(\w) & =-\frac{2g^{2}\gb}{N}\frac{Q(\w)+D}{\left(1+\w^{2}\right)^{2}}\frac{1-\bar{g}+\omega^{2}(1+\bar{g})}{(1-\bar{g})^{2}+\omega^{2}}
\end{align*}
From here we can solve for $Q(\w)$: 
\begin{align*}
Q(\w) & =\int\frac{d\w}{2\pi}Q(\t_{1})\exp(i\w\t_{1})\\
 & =g^{2}\l xx\r_{1}(\w)+I_{11}(\w)+I_{12}(\w)\\
 & =\frac{Q(\w)+D}{1+\w^{2}}\alpha(\w)\\
\Leftrightarrow Q(\w) & =D\frac{\alpha(\w)}{1-\alpha(\w)+\w^{2}}.
\end{align*}
with
\begin{align*}
\alpha(\omega) & =g^{2}\Big(1+\frac{1}{N}\frac{1}{(1-\gb)^{2}+\w^{2}}-\frac{1}{N}\frac{2\gb}{1+\w^{2}}\frac{1-\bar{g}+\omega^{2}(1+\bar{g})}{(1-\bar{g})^{2}+\omega^{2}}\Big)\\
 & =g^{2}\Big(1+\frac{1}{N}\frac{1}{(1-\gb)^{2}+\w^{2}}\left[1-2\gb(1+\gb-\frac{2}{1+\omega^{2}})\right]\Big).
\end{align*}

\end{document}